\documentclass[12pt,a4paper]{article}

\usepackage{amsmath}
\usepackage{amsfonts}
\usepackage{amssymb}
\usepackage{amsthm}

\input{epsf}             

\newcommand{\C}{\mathbb C}  
\newcommand{\SU}{{\rm SU}}
\newcommand{\U}{{\rm U}}

\newcommand{\Tr}{{\rm Tr}}
 
\theoremstyle{plain} 

\theoremstyle{definition}

\theoremstyle{remark}

\begin{document}

\title{Non-commutative geometry and the standard model vacuum}

\author{John W. Barrett and Rachel A. Dawe Martins
\thanks{Copyright \copyright\ John W. Barrett and Rachel A. Dawe 2006}
\\ \\
School of Mathematical Sciences\\
University of Nottingham\\
University Park\\
Nottingham NG7 2RD, UK\\
\\
john.barrett@nottingham.ac.uk\\
rachel.dawe@maths.nottingham.ac.uk}


\maketitle

\begin{abstract}   The space of Dirac operators for the Connes-Chamseddine spectral action for the standard model of particle physics coupled to gravity is studied. The model is extended by including right-handed neutrino states, and the $S^0$-reality axiom is not assumed. The possibility of allowing more general fluctuations than the inner fluctuations of the vacuum is 
proposed. The maximal case of all possible fluctuations is studied by considering the equations of motion for the vacuum. Whilst there are interesting non-trivial vacua with Majorana-like mass terms for the leptons, the conclusion is that the equations are too restrictive to allow solutions with the standard model mass matrix.
\end{abstract}

\section{Introduction}

\subsection{The non-commutative geometry of the standard model}
 Alain Connes discovered a geometric principle which unifies the metric tensor of general relativity with the classical fields of particle physics - both bosons and fermions - in one geometric structure \cite{C}. The principle is similar to Kaluza and Klein's idea of extending Einstein's space-time geometry to incorporate an `internal space' at each point in space-time so that the geometry of the extra dimensions is determined by the matter fields of particle physics \cite{OS}. Connes' principle is both a generalisation of this idea and a simplification. It is more general because the internal space is allowed to be a non-commutative `space', whereas it is at the same time simpler because the internal space is 0-dimensional. This is possible because in the non-commutative world 0-dimensional spaces have a much richer structure than in the commutative world of Kaluza and Klein, for whom a 0-dimensional space would have been merely an uninteresting discrete set of points. A non-commutative 0-dimensional space is characterised by a finite-dimensional matrix algebra, which turns out to be just what is required to produce non-abelian gauge fields. 
 
 In fact, non-commutative geometry explains the geometrical structure of the standard model of particle physics \cite{CR}.  One of the most striking features of this is the discovery that the Higgs field and the gauge bosons are both part of a connection on the total geometry - space-time plus internal space. The Higgs field is the part of the connection in the internal space `directions'. These directions are actually discrete, so one can think of the Higgs field as providing the parallel transport for hops from left- to right-handed particles, and vice-versa. This geometrical picture extends to a unified formula for the particle physics action, the famous quartic `mexican hat' potential for the Higgs being nothing other than part of the (quartic) Yang-Mills action for the connection on the total geometry. Another striking feature is the extension of the Yang-Mills gauge group to a matrix algebra. For example, $SU(3)$ becomes $M_3(\C)$, the algebra of all $3\times3$ matrices, which contains $\SU(3)$, but has a more restricted representation theory. The fact that the fermions fall into representations of the matrix algebra provides a deep explanation of the ad hoc pattern of charges which appears in the usual formulation of the standard model.  These insights are very striking and suggest that non-commutative geometry is an important part of particle physics.
 
 Connes' spectral triple formulation of a non-commutative geometry contains the following elements: A Hilbert space $\mathcal H$ with an involution $\gamma$ and an anti-linear involution $J$, a real *-algebra $\mathcal A$ of bounded operators in $\mathcal H$, and a Dirac operator $D$. Here we would like to give a brief description of these for the standard model; more details are given below in section \ref{moredetails}.

 The Hilbert space $\mathcal H$ is the space of classical fermion fields on the space-time manifold $M$. This is a finite number of Dirac spinors, one for each elementary fermion (left-handed leptons and quarks, right-handed-leptons and quarks, and their anti-particles). Thus $\mathcal H=\mathcal H_M\otimes \mathcal H_F$, where $\mathcal H_M$ is the Hilbert space of Dirac spinors on $M$ and $\mathcal H_F$ is the finite-dimensional Hilbert space with basis the elementary fermions.

The algebra $\mathcal A=\mathcal A_M\otimes\mathcal A_F$ is the algebra of coordinates on the total space, the product of the space-time with the internal space. Whilst the functions on spacetime $\mathcal A_M$ commute as usual, the internal space is non-commutative, so that its algebra of coordinates $\mathcal A_F$ is a matrix algebra. The operator $J$ is charge conjugation, while $\gamma$ is the chirality operator. Finally the Dirac operator contains all the bosonic fields (metric, gauge fields and Higgs), as well as the parameters for the fermion masses and their mixing angles. 
  
In \cite{SAP}, Connes and Chamseddine formulated a very simple formula for the action for this data, called the spectral action. The action is  
\begin{equation}\label{spaction} \Tr\left(\chi(D^2)\right) + \left(\psi,D\psi\right)\end{equation}
with $\psi\in {\cal H}$ the fermion field and $\chi$ a cut-off function on the spectrum of $D^2$ which interpolates between 1 below a very high-energy cut-off scale (possibly the Planck scale), and 0 above it.  Amazingly, there is a class of Dirac operators for which this unpacks to give the very long formula for the full standard model Lagrangian coupled to gravity. 

Impressive though this is, there are a number of issues that need to be resolved before it can have a greater impact in particle physics. In the Lagrangian of \cite{SAP}
 \begin{enumerate}
\item The space-time metric has Euclidean signature.
\item The fermions are quadrupled. For example there are separate Dirac spinors for the left-handed electron, right-handed electron and their antiparticles, whereas physically there should be only one \cite{fd1,fd2}.
\item The bare Weinberg angle is predicted as $\sin^2\theta=\frac38$. In \cite{SAP} this is assumed to be the high energy value, which changes under renormalisation. However renormalisation does not give the correct experimental value \cite{CIKS}.
\item The neutrinos are massless. Whilst this is correct for the standard model, observational evidence shows that neutrinos have mass, and so the non-commutative geometry should be extended to account for neutrino masses.
\end{enumerate}

In addition to this, there are some further theoretical puzzles about seemingly ad hoc features of the action. In contrast to the above, these points do not indicate problems with the physics of the model, but raise questions about the mathematical formulation and about the understanding of the standard model in terms of non-commutative geometry.

\begin{enumerate}
\item[5.] To obtain the standard model, it is necessary to remove a U(1) gauge field in $D$ by the `unimodularity constraint' \cite{C,LS,IKS}. It is known that this is equivalent to the requirement that the resulting quantum field theory is anomaly-free \cite{CH,ANO}, but the reason for this equivalence is mysterious.
\item[6.] To obtain the standard model, a particular vacuum Dirac operator $D_0$ is chosen. Are there any theoretical constraints on this choice, or are there many other (non-physical) variants of the standard model?
\item[7.] Is there an intrinsic definition (in terms of non-commutative geometry) of the set of Dirac operators that the action (\ref{spaction}) applies to?
\end{enumerate}

\subsection{Sets of Dirac operators}

Our main observation is that the theoretical puzzles are all questions about which is the set of Dirac operators to which the spectral action should be applied.  Our contribution is to study what happens when the set of Dirac operators is enlarged to the maximum extent allowed by the axioms.

Accordingly, we first explain the class of Dirac operators which appears in Connes' standard model. In Connes' construction, given a metric (and spin-structure) $g$ on $M$, a `vacuum' geometry is defined by the Dirac operator on $\cal H=\cal H_M\otimes \cal H_F$
\begin{equation}\label{vacuum}D_0=D_M\otimes I+ \gamma_M\otimes D_F,\end{equation}
where $D_M$ is the usual Dirac first-order differential operator of $g$ with the Levi-Civita connection, $\gamma_M$ is the chirality operator on $(M,g)$ (often denoted $\gamma_5$)
and $D_F$ is a certain matrix which encodes the vacuum Higgs fields and the fermion mass matrix. The details of $D_F$ are explained below in section \ref{moredetails}. Then the gauge and Higgs fields for this metric are obtained by the process of `internal fluctuations' determined by the 1-forms $A=A^*=\sum_i a_i[D_0,b_i]$ given a finite set of elements $a_i, b_i\in \cal A$.
The result is the set of Dirac operators\footnote{Note: in an arbitrary dimension the correct formula is $D_0+A+\epsilon'JAJ^{-1}$, where $\epsilon'=\pm1$ is determined by $JDJ^{-1}=\epsilon'D$; $\epsilon'=1$ in all even dimensions.}
\begin{equation}\label{innerfluctuations}{\mathcal D}^g=\{D_0+A+JAJ^{-1}\}.\end{equation}
However the unimodular condition has to be taken account of. This restricts to a smaller set of Dirac operators ${\mathcal D'}^g\subset{\mathcal D}^g$ obtained by removing a $\U(1)$ gauge field that has charge $+1$ for all quarks, $0$ for leptons and $-1$ for anti-quarks.
Finally, the set of all Dirac operators for the Connes-Chamseddine spectral action is the union of these for all metrics and spin structures,
$${\mathcal D_{CC}}=\bigcup_g {\mathcal D'}^g.$$ 

It appears that a similar construction could be carried out starting with any matrix $D_F$. The only difficulty would be deciding exactly how the unimodularity condition generalises; the most physical generalisation would be a condition that guarantees the absence of anomalies in the corresponding perturbative quantum field theory.

An answer to question 6 is provided by the observation that the vacuum Dirac operator should be a stationary point for the spectral action. Therefore there are equations which restrict $D_0$. These equations require that the Higgs fields in $D_F$ are actually at a stationary point of the Higgs potential. There are no constraints on the parameters of the fermion mass matrix.

From the point of view of non-commutative geometry, the Connes-Chamse\-ddine class of Dirac operators is very strange, as its description uses the underlying commutative description of the fields and requires an unexplained choice of $D_F$ which is rather special. From the point of view of physics one can describe this by saying that the Connes-Chamseddine class of Dirac operators treats the gravitational and the bosonic matter degrees of freedom differently; for the bosons the internal fluctuations are used but for the gravitational degrees of freedom, all possible fluctuations of the Dirac operator are used, the internal fluctuations being trivial. This undermines the idea of a geometric unification of matter and gravity.

A much more natural class operators (alluded to in \cite{C}) is the set $\mathcal D$ of all Dirac operators for the standard model $\mathcal A$ and $\mathcal H$. From a physical point of view, this class is almost certainly too big; however understanding the consequences of choosing $\mathcal D$ for the spectral action is a necessary first step towards investigating whether there is a natural class which is larger than $\mathcal D_{CC}$ but smaller than $\mathcal D$.

Enlarging the set of Dirac operators for the action has two effects. Firstly, some of the parameters in $D_0$ which were previously constants now become variables. Secondly, there are an equal number of new equations of motion which, in the generic situation, will therefore fix the vacuum values of this number of constants. So, for example, taking the extreme case where the set of Dirac operators is just one Dirac operator, $\{D_0\}$, then there are no equations of motion and so no constraints on any of the parameters in $D_F$. Then enlarging the class of Dirac operators to the set ${\mathcal D}^g$ given by the internal fluctuations of a vacuum $D_0$, results in promoting the Higgs vacuum parameters to fields. The corresponding new equations of motion, as noted above, demand that the vacuum values of these parameters lie at the stationary points of the Higgs potential, which is a significant constraint. The gauge fields have equations of motion of course, but these are all compatible with the vacuum values zero, and the unimodularity constraint does not affect the vacuum either. Looking at the class $\mathcal D_{CC}$, one has in addition the Einstein equations for the variation of the metric $g$. 

Enlarging the class further brings the prospect of further Higgs fields and further constraint equations. The danger is that the additional equations may rule out the standard model vacuum, or provide additional fields which contradict phenomenology. However if these dangers do not materialise, there is the major benefit that further equations will provide previously unknown relations between the parameters of the standard model in the fermion mass matrix, and in models which are extended to allow neutrino masses, may have predictive power in constraining the form of the neutrino sector. 

In this paper we consider the set $\mathcal D$ of all Dirac operators for the given $\mathcal H$ and $\mathcal A$ for the standard model, and also for the model where $\mathcal H$ is enlarged to allow a right-handed neutrino. This approach contrasts with that of \cite{Schucker et al}, where enlarging the set of Dirac operators was considered by enlarging the algebra $\cal A$ but staying with the class of inner fluctuations.
We assume the vacuum is of the form (\ref{vacuum}), with $g$ a flat metric. Thus we are essentially ignoring the Einstein equations, which of course would be important on a macroscopic scale (e.g. in cosmology), but not microscopically. We calculate the equations of motion for $D_F$ by requiring that it is stationary for all variations of the action in this class and classify the possible vacua under some simplifying assumptions. Finally the physical relevance of the equations and the vacua we have found is addressed in the concluding remarks.

  \section{Details of the standard model}\label{moredetails}
 The internal Hilbert space is
 \begin{equation}\label{hf}\mathcal{H}_F
= \mathcal{H}_L \oplus \mathcal{H}_R \oplus \mathcal{H}_L^c \oplus \mathcal{H}_R^c,\end{equation}
 where
\begin{displaymath}
   \mathcal{H}_L = ( \mathbb{C}^2 \otimes \mathbb{C}^n \otimes \mathbb{C}^3 )
   \oplus ( \mathbb{C}^2 \otimes \mathbb{C}^n )
\end{displaymath}

\begin{displaymath}
 \mathcal{H}_R = ( (\mathbb{C} \oplus \mathbb{C} ) \otimes \mathbb{C}^n \otimes \mathbb{C}^3 )
   \oplus ( \mathbb{C} \otimes \mathbb{C}^n )
\end{displaymath}
A basis of $\mathcal H_F$ is labelled by the elementary fermions and their antiparticles. The symbol $c$ is
used to indicate the subspace represented by the antiparticles, which duplicates the particle space. In either case of $\mathcal{H}_L$
and $\mathcal{H}_R$, the first direct summand is the quarks and the second, the leptons. 
The first factor in the tensor product is the down/up (or electron/neutrino) doublet, the second factor is the space of $n$ generations, and the third factor, for quarks, is colour. 

Since the fermions are left- or right-handed, $\mathcal{H_F}$ is $\mathbb{Z}/2$ graded   by the chirality operator \mbox{$\gamma_F$ = diag$(1,-1,1,-1)$}, using the decomposition (\ref{hf}).   
The right-handed neutrino $\nu_R$ does not occur in the standard model, but we include in as an extension of the standard model which allows neutrino masses.   Results for models without the right-handed neutrino can be easily obtained by setting the relevant matrix entries in $D_F$ to zero and dropping the equations obtained by varying them. In the following, explicit matrices are written and so we need a convention for the order of the rows and columns: the quark basis is $(d_L, u_L, d_R, u_R)$, each of which is reproduced in three colours, and a similar basis  of singlets for the leptons $(e_L, \nu_L, e_R, \nu_R)$. The antiparticle bases are correspondingly $(d_L^c, u_L^c, d_R^c, u_R^c)$ and $(e_L^c, \nu_L^c, e_R^c, \nu_R^c)$.

The Standard Model algebra is
\begin{equation*}
   \mathcal A_F = \mathbb{H} \oplus \mathbb{C} \oplus M_3(\mathbb{C})
\end{equation*}
where $\mathbb{H}$ are the quaternions.
The action of an element $(q,\lambda,m)$ of $\mathcal A_F$ is 
\begin{equation}\label{algaction}
\rho=
\left( \begin{array}{cccc}
\rho_L   &       0     &       0       &     0  \\
0        &    \rho_R   &       0       &     0  \\
0        &      0      &    \rho_L^c   &     0  \\
0        &      0      &       0       &  \rho_R^c
\end{array}   \right)
\end{equation}
where $\rho_L=q$ acting on isospin, $\rho_R= \Lambda =
\left(   \begin{array}{cc}
\bar\lambda &  0\\
0           &    \lambda
\end{array}  \right)
$
acting on the two isospin scalars, $(d_R,u_R)$ or $(e_R,\nu_R)$. 
The action on the antiparticles is
$\rho_L^c = \rho_R^c = m $ acting on the colour index for quarks, and   $\rho_L^c = \rho_R^c = \lambda$ for leptons. The action is the same for each generation, and we refer to the analogues of the down, up, electron and neutrino in the other generations by the same names.
 
A real spectral triple possesses a real structure given by an operator $J$
that takes particles into antiparticles and charge conjugation.  
\begin{displaymath}
   J \binom{ \psi_1}{\psi_2^c}=\binom{\bar\psi_2^c}{\bar\psi_1}
        \quad \in \mathcal{H}
\end{displaymath}
It turns the Hilbert space into a bimodule.
\begin{equation}    \label{bimodule}
\quad a\psi b=aJb^{\ast}J^{-1}\psi \end{equation} 
Note that \begin{equation}\label{notethat} [a,Jb^{\ast}J^{-1}]=0.
\end{equation} 
Also, $J$ commutes with $\gamma=\gamma_M\otimes\gamma_F$.

This data satisfies an axiom called Poincar\'e duality, which is the generalisation of the familiar Poincar\'e duality of manifolds to non-commutative geometry. Whilst this axiom is a natural generalisation from the mathematical point of view, its physical meaning for the internal space is unclear. Applied to the internal geometry, the axiom is satified for the standard model but not if there is the same number of left- and right-handed neutrinos \cite{Schelp}. However it is satisfied if one of the generations does not have a right-handed neutrino but the other two do. In the following we do not require that the Poincar\'e duality axiom holds.

So far, we have explained some of the axioms relating to the Hilbert space and the algebra for the Standard
Model finite triple. The final ingredient is the Dirac operator. A Dirac operator must
satisfy the first order condition:
\begin{equation}
   [[{D},a],Jb^*J^{-1}]=0
\end{equation}
for all $a,b\in\cal A$ in order that $D$ be a first order differential operator \cite{Dubois}. Note that due to (\ref{notethat}), this implies that also  $[[D,Jb^*J^{-1}],a]=0$.

 The choice made for $D_F$ in \cite{SAP} in order
that the Spectral action principle reproduces the Standard Model is:
\begin{equation}        \label{ad hoc choice}
D_F =
\left(   \begin{array}{cccc}
0         &         M^{\ast}    &   0        &   0\\
M         &          0          &   0        &   0\\
0         &          0          &   0        & M^T\\
0         &          0          &   \bar{M}  &  0
\end{array}  \right)
\end{equation}
where $M = Q\otimes 1_3 \oplus L$,  
\begin{displaymath}
Q^* =
\left(   \begin{array}{cc}
  k_d \phi_1         &     k_u \bar{\phi_2}  \\
 k_d \phi_2    &        -k_u \bar{\phi_1}
\end{array}  \right)
\end{displaymath}
\begin{displaymath}
L^* =
\left( \begin{array}{cc}
     k_e \phi_1    &     0  \\
   k_e {\phi_2} &  0
\end{array} \right)
\end{displaymath}
with
\begin{displaymath}
k_u =
\left(   \begin{array}{ccc}
m_u   &     0     &    0   \\
0     &    m_c    &    0   \\
0     &     0     &     m_t
\end{array}  \right)
\qquad k_d =
V_{CKM}
\left(   \begin{array}{ccc}
m_d   &     0     &    0   \\
0     &    m_s    &    0   \\
0     &     0     &     m_b
\end{array}  \right)
\end{displaymath}
\begin{displaymath}
k_e =
\left(   \begin{array}{ccc}
m_e   &     0     &    0   \\
0     &    m_{\mu}    &    0   \\
0     &     0     &     m_{\tau}
\end{array}  \right)
\end{displaymath}
$M^T$ denotes the transpose, $M^\ast$ denotes hermitian conjugate, $\bar M$ denotes the complex conjugate matrix,
$m_x$ are the Yukawa couplings of the elementary fermions, $V_{CKM}$ is the
Cabibbo-Kobayashi-Maskawa generation mixing matrix, and $(\phi_1, \phi_2)$ is the Higgs scalar doublet.

\section{The general $D_F$}\label{general D}

The problem of finding the vacua reduces in essence to 
 considering a single point of space time. We simplify the action formula (\ref{spaction}) by
removing all terms involving the space time
curvature, and all kinetic terms, and set the gauge fields to zero. This is equivalent to varying the Dirac operator over all operators of the form (\ref{vacuum}), with a fixed $D_M$. By inspecting the heat expansion detailed in
\cite{Gilkey} we find that the action we are looking for is:
\begin{equation}           \label{S}
  S = \Tr(D_F^4 - 2 D_F^2)
\end{equation}
This formula gives the Higgs potential for internal fluctuations of the Standard Model vacuum. However it applies to the wider class of operators $\mathcal D$.

In order to find the most general $D_F$ for the Standard Model internal space, we employ the  
constraints given by the axioms for a 0-dimensional noncommutative space. Only the axioms involving the Dirac operator are listed here.

\begin{description}
\item[I. Self-adjointness]\quad
  ${D_F} = {D_F}^{\ast}$
\item[II. {Reality}]\quad
$  [{D_F},J]=0$\end{description}
These first two imply that, splitting the Dirac operator into four blocks corresponding to the particle/antiparticle split of the basis,
\begin{equation}    \label{D with Z}
D_F=
\left(   \begin{array}{cc}
Y & Z\\
\bar{Z} & \bar{Y}
\end{array}  \right)
\end{equation}
where $Y=Y^{\ast}$ and $Z=Z^T$.

\begin{description}
\item[III. Orientability]\quad
  ${D_F} \gamma_F+\gamma_F{D_F} = 0$
\end{description}
implies that
\begin{equation}    \label{D_F}
D_F =
\left(   \begin{array}{cc}
Y   &    Z   \\
\bar{Z}   &  \bar{Y}
\end{array}  \right)
\quad = \quad
\left(   \begin{array}{cccc}
0         &       M^{\ast}     &     0      &    G   \\
M         &       0            &     G^T    &    0   \\
0         &   \bar{G}          &     0      &    M^T \\
G^{\ast}  &       0            &  \bar{M}   &    0
\end{array}   \right)
\end{equation}
using a further split of each block into the left/right subspaces.
In this formula $M$ and $G$ are general matrices with complex coefficients, $M$ giving a generalisation of the mass matrix of the standard model and $G$ having the interpretation of Majorana mass terms and other interactions.

If the vacuum Dirac operator generates Majorana masses then the model is no
longer the Standard Model but a modification of it. In this modified model, the generation of
Majorana masses involves symmetry breaking with new scalar fields, in addition to
the Higgs fields. This introduction of new fields is not necessarily undesirable
because any new physics of fermion masses must necessarily go beyond the Standard Model.

The remaining axiom is
\begin{description}
 \item[IV. First order condition]\quad
$ [[D_F,a],Jb^*J^{-1}]=0$ 
\end{description}
The effect of applying the first order condition to (\ref{D_F}) is to determine which
elements of $M$ and $G$ are non-zero. 

In \cite{Kraj}, Krajewski shows that for any finite dimensional spectral triple, the Dirac operator solves the
first order condition uniquely in the form of the sum:  $ D_F = D_L + D_R$ where
$D_R$ commutes with any element $b^o=Jb^*J^{-1}$ in the opposite algebra and  $D_L$ commutes
with any element in the algebra. See also \cite{PS}. Using these formulae
\begin{equation}       \label{1st Kraj}
   D_L a - a D_L=0,  \qquad   D_R b^o - b^o D_R = 0
\end{equation}
 and using the representation as given above  
\ref{algaction} we find that
\begin{equation}
    D_L =
\left(   \begin{array}{cccc}
0         &       0            &     0      &    0      \\
0         &       0            &     G^T    &    0   \\
0         &   \bar{G}          &     0      &    M^T \\
0         &       0            &  \bar{M}   &    0
\end{array}   \right)
\end{equation}
and
\begin{equation}        \label{DR}
D_R =
\left(   \begin{array}{cccc}
0         &       M^{\ast}     &     0      &    G   \\
M         &       0            &     0      &    0   \\
0         &       0            &     0      &    0  \\
G^{\ast}  &       0            &     0      &    0
\end{array}   \right)
\end{equation}
with
M=$Q\otimes1_3\oplus L$, a direct sum of a quark matrix $Q$ which commutes with colour, and a lepton matrix $L$. Both $Q$ and $L$ are arbitrary $4\times4$ complex matrices (for one generation). For 3 generations, these becomes arbitrary $12\times12$ matrices. The other matrix is in block form
\begin{equation}    \label{G}
G =
\left(   \begin{array}{cccc}
0         &       0            &     0      &    0   \\
0         &       U            &     0      &    N
\end{array}   \right),
\end{equation}
using the basis explained in section \ref{moredetails}, i.e., a map 
$(d^c_R, u^c_R,  e^c_R,\nu^c_R)\mapsto$\\$(d_L, u_L, e_L,\nu_L)$. The non-zero entries are the blocks $U$ and $N$. These are maps
$$U\colon u_R^c\mapsto(e_L,\nu_L)$$
and 
$$N\colon \nu_R^c\mapsto(e_L,\nu_L).$$
Their appearance results from the the fact that $u^c_R$, $\nu^c_R$,  and $e_L$ and $\nu_L$ are all in the same representation of the opposite algebra $\mathcal A^o$, multiplication by the complex number $\lambda$. The other entries of $G$ are zero because they are intertwining inequivalent representations. Note that the matrix $\bar G$ gives maps $u_R\mapsto(e^c_L,\nu^c_L)$ and $\nu_R\mapsto(e^c_L,\nu^c_L)$.

Explicitly, for one generation we use the matrices
\begin{equation}\label{QL}
Q =
\left( \begin{array}{cccc}
d     &     c     \\
b     &     a    \\
\end{array}   \right),\quad
L =
\left( \begin{array}{cccc}
      q     &    r\\
      s     &    t
\end{array}   \right)
\end{equation}
\begin{equation}\label{UN}
U =
   \left( \begin{array}{ccc}
x & u & g \\
y & v & h  
\end{array} \right),\quad
  N =   \left( \begin{array}{c}
 j\\
 l
\end{array} \right)                  
\end{equation}
$(x,u,g), (y,v,h)$ are three-dimensional colour vectors and $j,l$ are colour singlets.  

It is also convenient to split $Q$ and $L$ into smaller blocks corresponding to the gauge-invariant split of the right-handed fermions into down/up or electron/neutrino:
\begin{equation}\label{qlsplit}
Q =
\left( \begin{array}{cccc}
Q_d    \\
Q_u   \\
\end{array}   \right),\quad
L =
\left( \begin{array}{cccc}
      L_e\\
      L_\nu
\end{array}   \right)
\end{equation}
For example, for one generation this means that $L_e=\begin{pmatrix}q& r\end{pmatrix}$, etc.

The non-zero entries give rise to terms in the fermionic part of the action (\ref{spaction}) given, for one generation, by
\begin{equation*}   \bigl(\psi_L^c, \bar{G} \psi_R\bigr) =    \bigl(e_L^c, (\bar x,\bar u,\bar{g})\cdot u_R\bigr) + \bigl(\nu_L^c, (\bar y,\bar v,\bar{h})\cdot u_R\bigr)
+ \bigl(e_L^c, \bar{j} \nu_R\bigr) + \bigl(\nu_L^c, \bar{l} \nu_R\bigr)\end{equation*}
\begin{equation*} 
\bigl(\psi^c_R, G^{\ast} \psi_L\bigr) = \bigl(\nu_R^c,
\bar{j} e_L\bigr) + \bigl(\nu_R^c, \bar{l} \nu_L\bigr) + \bigl(u_R^c,  (\bar x,\bar u,\bar{g}) e_L\bigr) + \bigl(u_R^c, (\bar y,\bar v,\bar{h}) \nu_L\bigr)
\end{equation*}
plus the hermitian conjugate of each term, which are $(\psi_R, G^T \psi_L^c)$ and $(\psi_L, G
\psi_R^c)$ respectively.
  Thus the action contains, for example,
\begin{displaymath}      \label{neutrino terms}
(\nu_R^c, \bar{l}~ \nu_L) + (\nu_L^c,\bar{l}~ \nu_R) +  (\nu_L,l ~ \nu_R^c)+(\nu_R, l ~\nu_L^c).
\end{displaymath}
  This action is plausibly the Euclidean analogue of a Majorana mass term; but note that significant differences between the Euclidean and Minkowskian formulation mean that this is somewhat heuristic.
The fields $U$ have been studied before in the context of the non-commutative standard model and are called leptoquarks
\cite{leptoquarks}, while the $N$ are new fields which are of course absent if the right-handed neutrino is not included in the model.

Relatively recently there has been new experimental evidence for neutrinos being
massive. 
There are two possibilities for theoretical neutrino
mass generation, the Dirac mass and the Majorana mass term.
For a particle to have a Dirac mass, both chiralities must be
present, so this would require $\nu_R$ and thus a modification of the Standard Model. To justify
its existence of the $\nu_R$, there must be an explanation as to why it remains undetected; it must
either be extremely massive or sterile (not interacting). A Majorana mass term is possible without 
$\nu_R$, but it requires an $\SU(2)$ Higgs triplet; many models include $\nu_R$ and a combination of Majorana and Dirac mass terms to render $\nu_R$ very heavy whilst leaving $\nu_L$ relatively light.

\section{Equations of motion}

The overall aim is to see if the Standard Model vacuum $D_F$ can be found as a solution  of the
internal space  equations without making any assumptions other than the axioms
themselves, instead of making the choice (\ref{ad hoc choice}) motivated from laboratory evidence.
To this end, we need to vary \emph{all} the degrees of freedom of $D_F$ which means that the
Yukawa couplings, and generation mixing angles (plus all the other fluctuations) are no longer
viewed as constants but as dynamical variables. In other words, we let $\mathcal D=\{ D_F \}$ form the
configuration space of the theory and calculate the internal space equations of motion.
 
To calculate the internal space
  equations of motion, we minimise the action (\ref{S}) with respect to the degrees of freedom
of both $M$ and $G$. The action is:
\begin{multline}        S = - 2 D_F^2  +  (D_F)^4\\
=   -2 \big(  G^{\ast} G + M^{\ast} M \big)
  + (M^{\ast} M)^2  +  ( G^{\ast} G)^2
  + 2\big(  M^{\ast}MGG^{\ast} +  MM^{\ast}G^T\bar{G}   +   MG\bar{M} \bar{G}  \big)\\
=\Tr\Bigl(-(LL^*+3QQ^*+UU^*+NN^*)+\frac12((L^*L)^2+3(Q^*Q)^2+(UU^*+NN^*)^2)\\
+L^*L(UU^*+NN^*)+L_\nu L^*_\nu N^T\bar N+ (Q_uQ^*_u\otimes I_3)U^T\bar U\Bigr).
\end{multline}

The results for varying with respect to $Q$, $L$, $U$ and $N$ are respectively
\begin{equation}\label{qequation}
Q^*\left(-3I+3QQ^*+\left( \begin{array}{cccc}
0&0   \\
0&\Tr_\text{col}(U^T\bar U)   \\
\end{array}   \right)
\right)=0
\end{equation}
\begin{equation}\label{lequation}
-L^*+L^*LL^*+(UU^*+NN^*)L^*+\begin{pmatrix}0 &L_\nu^*N^T\bar N\end{pmatrix}+
\begin{pmatrix}0& N\bar L_\nu \bar N\end{pmatrix}=0
\end{equation}
\begin{equation}\label{uequation}
U^*(-I+UU^*+NN^*+L^*L)+(\bar Q_u Q^T_u\otimes I_3)U^*=0
\end{equation}
\begin{equation}\label{nequation}
-N^*+N^*UU^*+N^*NN^*+N^*L^*L+\bar L_\nu L^T_\nu N^*+ \bar L_\nu\bar N L_\nu=0
\end{equation}
In equation (\ref{qequation}), the matrix is split into blocks according to the down/up split of the basis, and $\Tr_\text{col}$ denotes the trace over colour degrees of freedom. Thus each matrix block has size $n\times n$, where $n$ is the number of generations. 

In the following we analyse the solutions of these equation in various special cases, and then make some remarks about the general case.

\section{Solutions with $G=0$}
With $U=0$ and $N=0$, the equations of motion reduce to                                                                             
\begin{eqnarray}
Q^{\ast}(QQ^{\ast}-I)=0\label{pi2}\\
 L^{\ast}(LL^{\ast}-I)=0
\end{eqnarray}
 By multiplying (\ref{pi2}) on the left by
$Q$, we see that $QQ^{\ast}$ is a self-adjoint projection, and by multiplying the conjugate on the left by $Q^{\ast}$
that $Q^{\ast}Q$ is also a projector. 
Therefore, by definition, $Q\in M_n(\mathbb{C})$ is a partial isometry. Another characterisation of a partial isometry is that it is a projector multiplied by a unitary matrix. Obviously the same conclusion applies for $L$. 

These equations are the same as the ones obtained from assuming the additional `$S^0$-reality' axiom \cite{CR}, which has the effect that $U$ and $N$ are equal to zero in the action.  

The Standard Model vacuum (\ref{ad hoc choice}) is clearly not a solution of our  equations. To see this, we
write down $MM^*$

\begin{equation}              \label{M squared}
\left( \begin{array}{ccc}
k_d^*k_d ( \vert \phi_1 \vert^2 + \vert \phi_2 \vert^2) & 0 & 0\\
0 & k_u^*k_u ( \vert \phi_1 \vert^2 + \vert \phi_2 \vert^2) & 0\\
0  &  0  & k_e^*k_e ( \vert \phi_1 \vert^2 + \vert \phi_2 \vert^2) \\
0 & 0 & 0
\end{array} \right)
\end{equation}
so the vacua satisfying our equations of motion have  degenerate fermion
masses that are either 0 or 1 (times a constant, which has been omitted from the action). We note that in the case in which $\nu_R$ is absent, the left-handed neutrino is necessarily massless. This is a consequence of the fact that $L$ is not a square matrix in that case.

The impact of the following sections is to explore the way in which lifting the assumption of $S^0$-reality allows further vacua with $U$ or $N$ not equal to zero.

\section{Solutions with $G\ne0$}

\subsection{General}\label{gnonzero}
Throughout we assume that the quark mass matrix $Q$ is non-degenerate. This means that $Q^*$ can be cancelled from (\ref{qequation}) to give
\begin{equation}\label{newqequation}
-3I+3QQ^*+\left( \begin{array}{cccc}
0&0   \\
0&\Tr_\text{col}(U^T\bar U)   \\
\end{array}   \right)
=0.
\end{equation}
This equation can be solved for $QQ^*$, and thus for $Q$, up to multiplication on the right by a unitary operator, which is a symmetry. One very important feature (not always shared by the degenerate case) is that $QQ^*$ is block diagonal. This is an important feature of the standard model vacuum. Indeed, using the split (\ref{qlsplit}), the equation becomes the three equations
\begin{eqnarray*}
Q_dQ_d^*=I\\
Q_dQ_u^*=0\\
Q_uQ_u^*=I-\frac13\Tr_\text{col}\left(U^T\bar U\right)
\end{eqnarray*}
which shows that the down-quark masses are all equal to 1, the block-diagonal feature, and that the up-quark masses are split by a non-zero $U$. It is worth noting that the condition 
$Q_dQ_u^*=0$ fixes the form of $Q_u$ in the standard model vacuum (\ref{ad hoc choice}) if $Q_d$ is given the correct form. However there is no equation which constrains $Q_d$ to have the special form of (\ref{ad hoc choice}).

\subsection{One generation}    \label{single generation}

For one generation of fermions, we use the explicit matrices of (\ref{QL}) and (\ref{UN}).
The equations of motion are written out in the appendix. Equation (\ref{qequation}) becomes (\ref{ac}) to (\ref{dc}), (\ref{lequation}) becomes (\ref{qc}) to (\ref{tc}), (\ref{uequation}) becomes (\ref{gc}) to (\ref{yc}), and (\ref{nequation}) becomes (\ref{jc}) and  (\ref{lc}).

The equation (\ref{newqequation}) becomes the following
\begin{eqnarray}   \label{a to d colour}
\vert c \vert^2
+ \vert d \vert^2 - 1 = 0\\
 a \bar{c} + \bar{b}d = 0 \label{diageqn}\\
3 \vert a \vert^2 +  3 \vert b \vert^2 \quad + \quad  \vert g \vert^2 + \vert u \vert^2 + \vert x
\vert^2 \quad + \quad \vert h \vert^2 + \vert v \vert^2 + \vert y \vert^2 - 3 = 0\label{newaequation} 
\end{eqnarray}
As above, (\ref{diageqn}) implies that $QQ^*$ is a diagonal matrix.

\subsubsection{Solutions with $U=0$, $N\ne0$}

In this case, $Q$ decouples from the other fields and its equation reduces to (\ref{pi2}). Thus $Q$ is a unitary matrix and the quark masses are all equal to 1. To present the solutions to these equations we have used the symmetry afforded by a $2\times2$ unitary matrix acting on $e_L$ and $\nu_L$ to simplify one of the vectors $(q,r)$, $(s,t)$ or $(j,l)$. 

The possible solutions are all equivalent to
\begin{enumerate}
\item[(i)] $q=1$, $s=t=r=0$, $j=0$, $|l|=1$
\item[(ii)] $q=r=s=0$, $|t|^2 + |j|^2 = 1$, $l=0$
\item[(iii)] $q=1$, $s=r=0$, $|t|=1/2$, $j=0$, $|l|=1/2$
\item[(iv)]  $q=0$, $s=r=0$, $|t|=1/2$, $j=0$,$|l|=1/2$
\item[(v)] $r=s=t=0$, $|q|^2+|j|^2=1$, $l=0$.
\end{enumerate}

The proof that these are the only solutions is to consider combining the equations $N^*\text{(\ref{lequation})}-\text{(\ref{nequation})}L^*$. This leads to a set of algebraic conditions which reduce to the given solutions when substituted into the full equations of motion.

\subsubsection{Solutions with $U\ne0$}

We have been unable to find any explicit one-generation solutions with non-zero leptoquarks $U$. For the case where the quark and electron masses in the mass matrix $M$ are non-zero, we can prove that there are no such solutions. This has the corollary that there are no solutions where the matrix $M$ takes the form of a one-generation version of the standard model vacuum (\ref{ad hoc choice}).

The argument is as follows. Using a symmetry as above, we may assume that $r=0$ and $q\ne0$. Combining (\ref{rc}), (\ref{qc}) and (\ref{gc}) shows that $g=0$, since $|a|^2+|b|^2\ne0$. Similarly, $u=x=0$. Then (\ref{hc}) and (\ref{newaequation})
implies either $h=0$ or $\frac23(|v|^2+|h|^2+|y|^2) + |l|^2 + |t|^2 =0$, which also implies $h=0$. A similar argument shows $v=y=0$. Hence $U=0$.

If additionally, $N=0$, then the same conclusion also holds if the electron mass $q=0$ (see section \ref{3g} below). 

It is worth noting that the solutions considered here in the case that $L_\nu=0$ are the same as for the system obtained by omitting the right-handed neutrino $\nu_R$ from the action.

\subsection{Three generations}\label{3g}

For three generations we do not have complete results but note that the equations are of course solved by aggregating three one-generation solutions. In addition we
outline some general features of the solutions. 
We continue to assume that the quark masses are all non-zero; thus the remarks of section \ref{gnonzero} continue to hold. A general feature of the solutions to the equations can be found by taking the combination of equations $U^*\text{(\ref{lequation})}-\text{(\ref{uequation})}L^*$. This gives the equation
$$L_e U=0.$$
In the case where $N=0$, it also shows that 
$$L_\nu U=0$$
and equation (\ref{uequation}) reduces to
$$\left(U^*U-\frac13\Tr_\text{col}\left(U^*U\right)\otimes I_3\right)U^*=0$$
from which follows that 
$$U^*U=C\otimes I_3$$
for some $3\times3$ matrix $C$. Since the rank of $U^*U$ is at most 6, $C$ has rank at most 2.
As a consequence, only 2 out of the 3 quarks up, charm and top are given
different masses from that of the 3 down-family quarks.

The corresponding argument for 1 generation shows that the rank of $U^*U$ is at most 2, and hence $C$ must be zero. This agrees with the result found explicitly in the previous section, and holds even if $q=0$.

\section{Conclusions}

We have attempted to understand the standard model vacuum from a fundamental point of view involving non-commutative geometry. In particular, the question is: why does nature pick one particular vacuum geometry, i.e., one particular set of parameters in the fermion mass matrix? To attack this question, we investigated the simplest possible Ansatz for the set of Dirac operators in the action which is to promote all the degrees of freedom of the internal geometry, including the mass parameters, to be dynamical fields. This gives additional equations of motion which complement the usual Einstein and matter equations. The conclusion is that these equations exclude the standard model vacuum. Therefore there is something unexplained about either the physics of the standard model or the geometry of spectral triples. This is our overall conclusion.

The main problem is that the simplest solutions involve an unwanted degeneracy in the masses of the fermions. However there are some quite complicated vacua in which this problem is partially alleviated. 
Allowing an extension of the model to include a right-handed neutrino lifts the degeneracy of the lepton masses via equation (\ref{lequation}) in some solutions. 
The potential occurence of leptoquark fields lifts the degeneracy of the up-quark masses to some extent, but the degeneracy of the down quarks remains. In practice it seems to be hard to find vacua with non-zero values of the leptoquark field. With one generation and reasonable assumptions, the leptoquarks are always equal to zero. This is in a sense reassuring because the leptoquarks would break colour symmetry, but it does not help with the problem of quark mass degeneracy. With three generations, leptoquark fields are possible but we have found solutions affecting at most two out of the three generations, thus giving some sort of consistency with the one-generation result.

Some features of the very special standard model vacuum are automatically incorporated, whereas others are not. 
In particular there are typically many Higgs fields and there appears to be no constraint which forces the Higgs for the leptons and the quarks to be the same field. 
One can see this in our results for one generation. 
Since it is not possible to diagonalise both $Q$ and $L$ with a single
unitary transformation, the parameters $q$, $r$, $a$ and $b$ are independent, providing two Higgs doublets.

The extension to include the right-handed neutrino also introduces potential Majorana mass terms for the leptons. The explicit solutions we found allow (i) a Dirac mass for the electron and a Majorana mass for the neutrino, (ii) a Dirac mass for the neutrino and a mixing between $\nu_R$ and $e_L$, (iii) a Dirac mass for the electron and a neutrino with both Dirac and Majorana mass terms, (iv) a massless electron with a neutrino with both Dirac and Majorana mass terms, and (v) a Dirac mass for the electron and a massless neutrino with a mixing between $\nu_R$ and $e_L$. However it is worth emphasising that there are differences between the Euclidean and Lorenzian formulations for fermions which make it a difficult to draw conclusions from this for the physical Lorentzian case. In spite of this our overall conclusion that the masses are too degenerate stands, and this points to the need for modifications to the formalism if the overall objectives are to be retained. In future this could possibly be carried out by adding more constraints to the space of Dirac operators or by additional terms to the spectral action.

 \subsubsection*{Acknowledgement} R.A.D.M. thanks EPSRC for financial support.

\section{Appendix}

The equations of motion for one generation and three coloured quarks.

\begin{equation}        \label{ac}
\bar{a}( 3 \vert a \vert^2 + 3 \vert c \vert^2 + 3 \vert b \vert^2 + \vert g \vert^2 +   \vert u
\vert^2 + \vert x \vert^2  + \vert h \vert^2 +\vert v \vert^2  + \vert y \vert^2  -3) +
3 \bar{c}\bar{b}d =0 \end{equation}

\begin{equation}        \label{bc}
\bar{b}( 3 \vert b \vert^2 + 3 \vert d \vert^2 + \vert g \vert^2 + \vert u \vert^2 +\vert x \vert^2
+\vert h \vert^2 +\vert v \vert^2 +\vert y \vert^2 + 3 \vert a \vert^2  -3) + 3 \bar{a}\bar{d}c =0
\end{equation}

\begin{equation}        \label{cc}
\bar{c}( \vert a \vert^2 + \vert c \vert^2 + \vert d \vert^2  -1) + \bar{a}\bar{d}b =0
\end{equation}

\begin{equation}        \label{dc}
\bar{d}( \vert b \vert^2 + \vert d \vert^2 + \vert c \vert^2  -1) + \bar{c}\bar{b}a =0
\end{equation}

\begin{equation}        \label{qc}
\bar{q}( \vert q \vert^2 + \vert s \vert^2 + \vert r \vert^2 + \vert g \vert^2 +\vert u \vert^2 +\vert x \vert^2 + \vert j
\vert^2 -1) + \bar{r}(\bar{h}g +\bar{v}u   +\bar{y}x+ \bar{l}j + \bar{s}t) =0
\end{equation}

\begin{equation}        \label{rc}
\bar{r}( \vert r \vert^2 + \vert t \vert^2 + \vert q \vert^2 + \vert h \vert^2 +\vert v \vert^2 +\vert y \vert^2   + \vert l \vert^2 -1) + \bar{q}(\bar{g}h +\bar{u}v  +\bar{x}y+ \bar{j}l + \bar{t}s) =0
\end{equation}

\begin{equation}        \label{sc}
\bar{s}( 3\vert j \vert^2 + \vert q \vert^2 + \vert s \vert^2 +\vert t \vert^2 + \vert g
\vert^2 +\vert u \vert^2  +\vert x \vert^2 + \vert l \vert^2 -1) + \bar{t}(\bar{h}g +\bar{v}u  +\bar{y}x+ 2\bar{l}j+
\bar{q}r) =  0  \end{equation}

\begin{equation}        \label{tc}
\bar{t}( 3\vert l \vert^2 + \vert r \vert^2 + \vert t \vert^2 + \vert s \vert^2 + \vert h
\vert^2 +\vert v \vert^2  + \vert y \vert^2  + \vert j \vert^2 -1) + \bar{s}(\bar{g}h +  \bar{u}v  +\bar{x}y+ 2\bar{j}l +
\bar{r}q)  =   0  \end{equation}

\begin{equation}        \label{gc}
\bar{g}( \vert g \vert^2 + \vert u \vert^2 + \vert x \vert^2 + \vert h \vert^2 + \vert j \vert^2 + \vert q \vert^2 + \vert s \vert^2 + \vert a \vert^2 + \vert b \vert^2 -1) + \bar{h}(\bar{r}q +   \bar{x}y  +\bar{u}v+ \bar{t}s +
\bar{j}l) =0
\end{equation}

\begin{equation}        \label{hc}
\bar{h}( \vert g \vert^2 + \vert v \vert^2 +\vert h \vert^2 + \vert y \vert^2 + \vert l \vert^2 + \vert r \vert^2 + \vert t
\vert^2 + \vert a \vert^2 + \vert b \vert^2 -1) + \bar{g}(\bar{q}r +\bar{y}x  +\bar{v}u  + \bar{s}t + \bar{l}j) =0
\end{equation}

\begin{equation}        \label{uc}
\bar{u}( \vert u \vert^2 +\vert g \vert^2 +\vert x \vert^2 + \vert v \vert^2 + \vert j \vert^2 + \vert q \vert^2 + \vert s
\vert^2 + \vert a \vert^2 + \vert b \vert^2 -1) + \bar{v}(\bar{r}q +   \bar{x}y  +\bar{g}h+ \bar{t}s +
\bar{j}l) =0
\end{equation}

\begin{equation}        \label{vc}
\bar{v}( \vert u \vert^2 + \vert v \vert^2 +\vert h \vert^2 + \vert y \vert^2 + \vert l \vert^2 + \vert r \vert^2 + \vert t
\vert^2 + \vert a \vert^2 + \vert b \vert^2 -1) + \bar{u}(\bar{q}r +\bar{y}x  +\bar{h}g  + \bar{s}t + \bar{l}j) =0
\end{equation}

\begin{equation}        \label{xc}
\bar{x}( \vert u \vert^2 +\vert g \vert^2 +\vert x \vert^2 + \vert y \vert^2 + \vert j \vert^2 + \vert q \vert^2 + \vert s
\vert^2 + \vert a \vert^2 + \vert b \vert^2 -1) + \bar{y}(\bar{r}q +   \bar{u}v  +\bar{g}h+ \bar{t}s +
\bar{j}l) =0
\end{equation}

\begin{equation}        \label{yc}
\bar{y}( \vert x \vert^2 + \vert v \vert^2 +\vert h \vert^2 + \vert y \vert^2 + \vert l \vert^2 + \vert r \vert^2 + \vert t
\vert^2 + \vert a \vert^2 + \vert b \vert^2 -1) + \bar{x}(\bar{q}r +\bar{v}u  +\bar{h}g  + \bar{s}t + \bar{l}j) =0
\end{equation}

\begin{equation}        \label{jc}
\bar{j}( 3\vert s \vert^2 + \vert j \vert^2 + \vert g \vert^2 +\vert u \vert^2 +\vert x \vert^2 +  \vert l \vert^2 + \vert t
\vert^2  + \vert q \vert^2 -1) + \bar{l}(\bar{r}q + 2\bar{t}s +
 \bar{g}h   + \bar{u}v + \bar{x}y )=0  \end{equation}

\begin{equation}        \label{lc}
\bar{l}( 3\vert t \vert^2 + \vert h \vert^2 + \vert v \vert^2 +\vert y \vert^2 + \vert j \vert^2 +\vert l \vert^2 + \vert r
\vert^2  + \vert s \vert^2 -1) + \bar{j}(\bar{q}r + 2\bar{s}t +
\bar{h}g  + \bar{v}u + \bar{y}x  )=0  \end{equation}


\end{document}